\documentclass[10pt,a4paper,aps,pra,twocolumn,nobibnotes,pra,showpacs,superscriptaddress,nofootinbib]{revtex4-2}

\usepackage{graphicx}
\usepackage{dcolumn}
\usepackage{bm}
\usepackage{amsmath}
\usepackage{bbm}
\usepackage{color}

\usepackage{hyperref}
\usepackage[normalem]{ulem}

\begin{document}

\preprint{APS/123-QED}

\title{Quantum Crossovers Revealed by Local Measurements}

\author{A. C. S. Costa}
\affiliation{Departamento de F\'{i}sica, Universidade Federal do Paran\'{a}, Caixa Postal 19044, 81531-980, Curitiba, Paran\'{a}, Brazil.}
\author{E. C. Diniz}
 \affiliation{Universidade do Estado de Mato Grosso, Campus Tangará da Serra, Rodovia MT 358, Km 07 (s/n), Jardim Aeroporto, Tangará da Serra, Mato Grosso, Brazil.}
\author{O. P. de S\'a Neto}%
 \email{olimpiopereira@phb.uespi.br}
\affiliation{Coordena\c{c}\~ao de Ci\^encia da Computa\c{c}\~ao, Universidade Estadual do Piau\'i, 64202-220, Parna\'iba, Piau\'i, Brazil.
}%

\date{\today}

\begin{abstract}
Quantum crossover phenomena play a central role in few-body open quantum systems, yet their identification often relies on global or model-dependent indicators. In this work, we demonstrate that crossovers can be robustly characterized through purely local measurements, establishing a direct connection between local quantum Fisher information and the onset of crossover behavior. We further demonstrate that quantum obesity does not, in general, generalize the quantum steering ellipsoid volume as a universal indicator of crossover. Instead, we identify regimes in which the ellipsoid volume remains insensitive to the transition, while the relevant signatures are encoded in the behavior of the local Bloch vector. These results reveal a geometric distinction between local and global indicators of crossovers.
\end{abstract}

\maketitle

\section{Introduction} 

Quantum phase transitions are a cornerstone of condensed matter physics and quantum information, marking qualitative changes in the ground state as control parameters are varied at zero temperature~\cite{vojta2003}. In few-body systems, however, where the thermodynamic limit is absent, such transitions do not exhibit true nonanalytic behavior. Instead, they appear as quantum crossovers, characterized by rapid yet continuous changes in the system’s physical properties.

A particularly rich and subtle phenomenon arises in such open systems: the quantum crossover~\cite{heiss1990avoided}, a continuous but sharp change in the behavior of quantum observables and correlation measures as system parameters are tuned. Unlike traditional quantum phase transitions, these few-body crossovers occur without requiring a large number of degrees of freedom and are accessible within current experimental capabilities. To probe these crossovers, the quantum Fisher information (QFI) has emerged as a powerful tool~\cite{sun2010fisher,ying2024quantum,poulsen2022quantum,tiwari2025quantum}. It not only quantifies the sensitivity of a quantum state to small perturbations, making it central to quantum metrology, but also acts as a multipartite entanglement witness~\cite{saleem2025quantum} and reveals the onset of criticality~\cite{hu2025digital}. Importantly, in systems of a few qubits, the QFI depends on the choice of the measured subsystem, providing insight into localized signatures of the crossover.

A complementary geometric viewpoint is offered by the framework of quantum steering ellipsoids (QSEs)~\cite{du2021,liu2025quantum,uola2020,jevtic2014,du2021,milne2014}, which characterize the set of states in which one part of a bipartite system can steer the other via local measurements. Based on this, the concept of quantum obesity (QO) has recently been introduced~\cite{milne2014quantum,rosario2023swapping,rosario2024} as a novel quantifier of correlation.  
Compared to other measures of correlations, QO’s analytic tractability is an advantage, especially in studies of dynamical behavior and quantum criticality. Notably, QO also serves as a witness of entanglement and has been shown to carry clear signatures of crossovers in quantum systems~\cite{rosario2024}. 

Here, we argue that a unified and complete characterization of these complementary signatures naturally calls for a tripartite setting in which distinct couplings act as signal and probe channels. This architecture enables controlled access to local responses, bipartite observables, and multipartite entanglement within a single theoretical and experimental framework. In this work, we introduce and analyze a minimal tripartite model with tunable couplings, establishing it as an effective platform for the unified investigation of quantum crossovers in few-body systems. Additionally, the experimental implementation of this system and the Hamiltonian under analysis are provided in the Supplementary information, offering a concrete blueprint for demonstrating crossover indicators in systems with tunable coupling and providing an experimentally relevant framework that supports our theoretical findings. Furthermore, local filtering operations can amplify these signatures near the crossover point, offering a method to enhance the detectability of critical behavior.

\begin{figure}
\begin{center}
{\bf OBSERVATION 1}
\includegraphics[scale=0.12,angle=0]{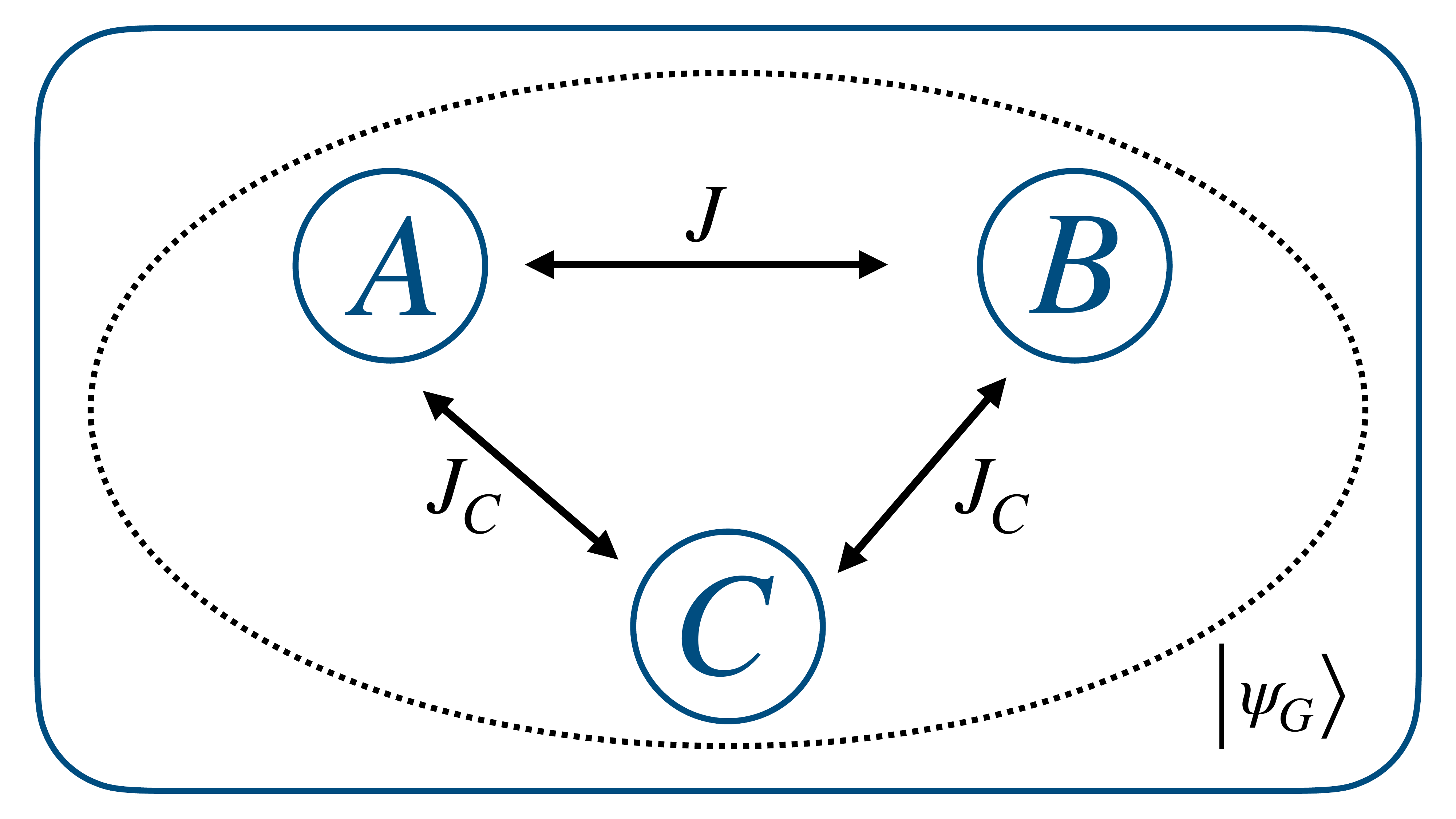}
{\bf OBSERVATION 2}
\includegraphics[scale=0.12,angle=0]{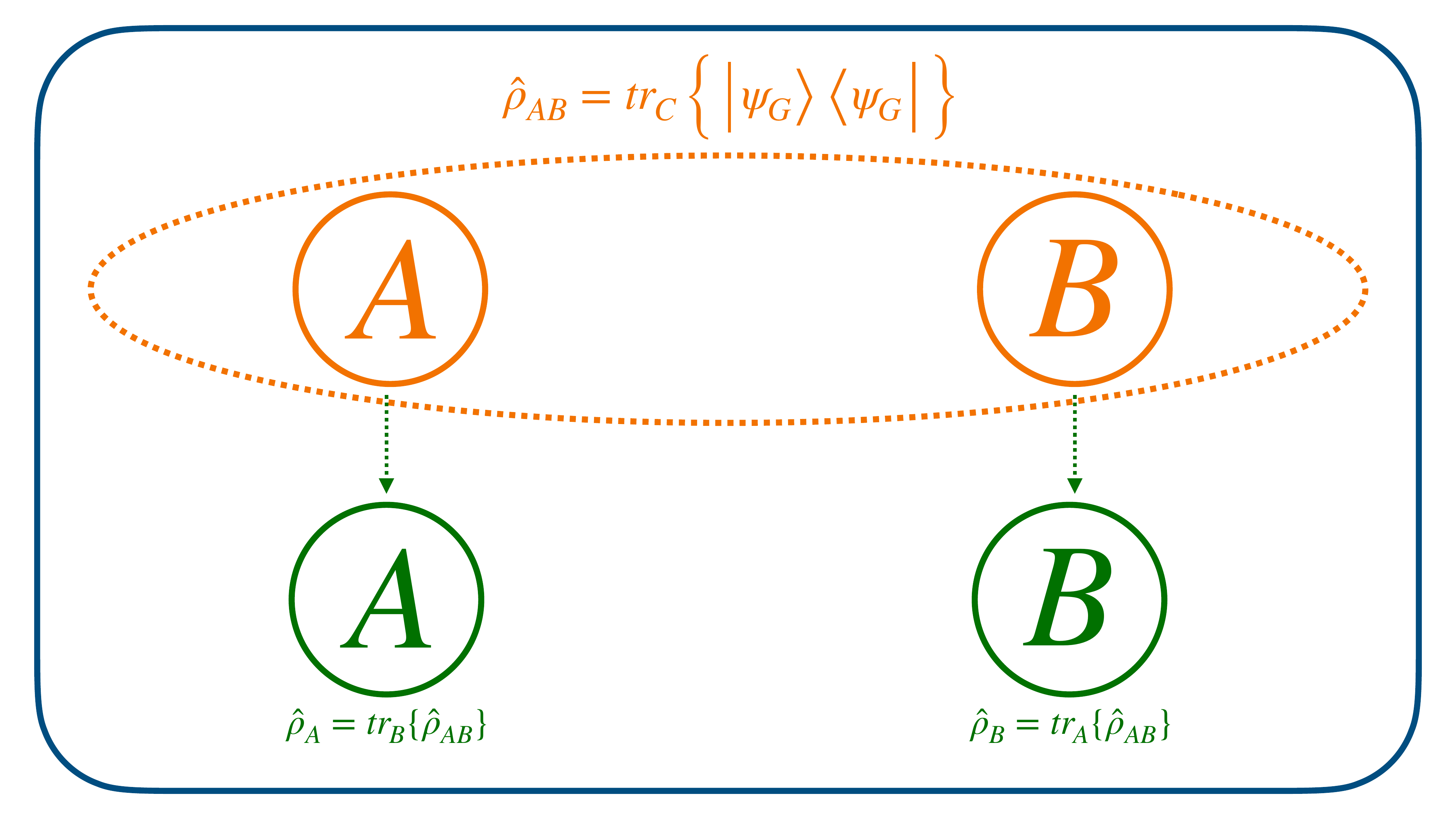}
{\bf OBSERVATION 3}
\includegraphics[scale=0.12,angle=0]{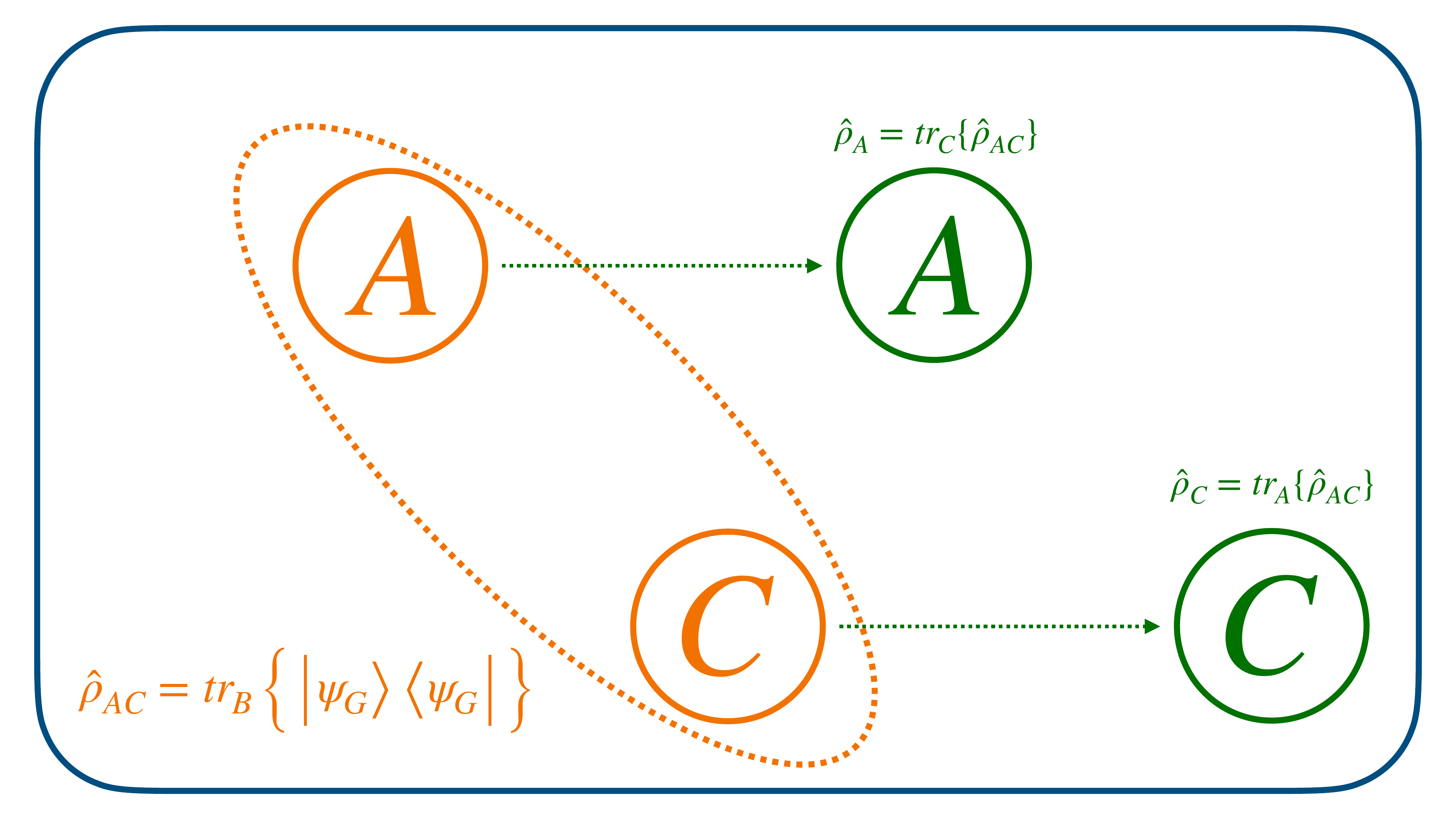}
\end{center}
\caption{Schematic representation of the tripartite system composed of three qubits $A$, $B$, and $C$.
Qubits $A$ and $B$ are identical and interact via a symmetric signal coupling of strength $J$.
Both qubits are independently coupled to the third qubit $C$ through probe couplings of strength $J_C$.
The figure illustrates the different observation schemes considered in this work:
(i) the full tripartite system,
(ii) the bipartition $AB$ obtained by tracing out qubit $C$, and
(iii) the bipartitions $AC$ and $BC$, obtained by tracing out one of the identical qubits.
These configurations allow the analysis of local measurements, bipartite correlations,
and symmetry-induced effects within a unified framework.} 
\label{FigABC}
\end{figure}

\section{Partition and Bipartition}

The central objective of this section* is to analyze the nature of local measurements at the level of single–qubit partitions and the emergence of nonclassical properties in bipartitions, as dictated by the symmetry and antisymmetry of the coupling structure. In particular, we are interested in understanding how symmetric signal couplings and probe–like interactions manifest differently in local observables and in bipartite correlations. The most natural framework to address this question is a tripartite system composed of two identical qubits, forming a symmetric partition connected by a signal coupling, which are both coupled to a third qubit through probe couplings. This architecture allows us to clearly separate local measurement effects from genuinely nonclassical bipartite properties. A schematic representation of the system and the considered partitions and bipartitions is shown in Fig.~\ref{FigABC}.

Considering a model with the effective Hamiltonian, we can write it as~\cite{fischbach2015} 
\begin{eqnarray}
\hat{H}_{S} &=& \hat{H}_{0} + \hat{H}_{I},
\label{HFull0}
\end{eqnarray}
where
$\hat{H} _{0} = \frac{\hbar\omega_{0} }{2}\left ( \hat{\sigma}_{z}^{A} +  \hat{\sigma}_{z}^{B}\right ) + \frac{\hbar\omega_{C} }{2} \hat{\sigma}_{z}^{C}$,
is the free Hamiltonian of the system, composed of the three qubits $A$, $B$, and $C$. We assume that qubits $A$ and $B$ are identical and have the same eigenfrequency $\omega_{A}=\omega_{B}=\omega_{0}$. The eigenfrequency $\omega_{C}$ is related to the third qubit $C$. The  Hamiltonian that describes the interaction between the qubits is
$\hat{H} _{I} = \hbar J \hat{\sigma}_{x}^{A} \hat{\sigma}_{x}^{B} + \hbar J_{C} \left ( \hat{\sigma}_{x}^{A} \hat{\sigma}_{x}^{C} + \hat{\sigma}_{x}^{B} \hat{\sigma}_{x}^{C} \right )$,
where $J$ is the coupling strength between qubits $A$ and $B$, and $J_{C}$ is the coupling strength between the identical qubits and qubit $C$. The experimental implementation of this system and the Hamiltonian described in Eq.~\eqref{HFull0} are provided in the Supplementary information.

The dissipative dynamics of the system-bath interaction at temperature $T = 0$ K, is given by the microscopic master equation (MME) ~\cite{rivas2012, breuer2007,diniz2021quantum}  
\begin{equation}
\label{eq:PMEspin0}
\frac{\partial\hat{\rho}_{S}(t)}{\partial t} = \mathcal{L}[\hat{\rho}_{S}(t)] = -\frac{i}{\hbar} [\hat{H_S},\hat{\rho}_{S}(t)] + \mathcal{D}\left[\hat{\rho}_{S}(t)\right],
\end{equation}
where $\hat{\rho}_S(t)$ is the time-dependent density matrix of the system and $\mathcal{L}$ is the Liouvillian superoperator describing the losses. The dissipator is given by 
{\small
\begin{equation}
\mathcal{D}\left[\hat{\rho}_{S}(t) \right] = \sum_{j=A,B,C}\sum_{\omega }J\left(
\omega \right)\left[ \hat{A}_{j}(\omega )\hat{\rho} _{S}\hat{A}_{j}^{\dagger }(\omega )-\frac{ \left\{ \hat{A}_{j}^{\dagger }(\omega )\hat{A}_{j}(\omega ),\hat{\rho} _{S}\right\} }{2}
\right],
\end{equation}}
where $J(\omega )=\mu_j\omega$ is the spectral Ohmic nature of the environment, and the operator
$\hat{A}_{j}(\omega) = \sum_{\epsilon_k-\epsilon_i=\omega}|\epsilon _i\rangle\langle\epsilon _i| \hat{A}_{j}|\epsilon _k\rangle\langle \epsilon_k|$
describes the coupling of the system with the environment, where $\omega$ is the energy difference relating the eigenstates $|\epsilon _i\rangle$ and $|\epsilon _k\rangle$ of the full Hamiltonian (\ref{HFull0}). For the heat bath, we have spontaneous decay, which can be described by using $\hat{A}_{j} = \hat{\sigma}_x^{j}$. 
The steady-state solution of the MME~\eqref{eq:PMEspin0} is the Gibbs state  $\hat{\rho}_{SS}(\infty)=\exp\left(-\beta \hat{H}_{S} \right)/\textrm{Tr}\{\exp\left(-\beta \hat{H}_{S}\right)\}$~\cite{diniz2021quantum,valente2020frustration,kuwahara2020clustering,chen2025efficient}. 

Our analysis is based on the knowledge of the global ground state, which reduces to the steady state of the system. The ground state for the full Hamiltonian (\ref{HFull0}) can be analytically written as 
\begin{equation}
\label{eq:psiG}
    \left| \psi_G \right\rangle = \frac{1}{\sqrt{\mathcal{N}}}
    \begin{pmatrix}
 0, &
- \alpha, &
 - \beta, &
  0, &
 - \beta, &
   0, &
   0, &
   1
\end{pmatrix}^{T},
\end{equation}
where $\mathcal{N}$ is the normalisation, and $\alpha$ and $\beta$ are explicitly described in the Methods, and depend on the system parameters.

From a structural point of view, the bipartite states obtained by tracing out one subsystem can be interpreted as effective purifications of the remaining partition. In this sense, each bipartition encodes, in a larger Hilbert space, information about the loss of purity induced by the partial trace.
While the reduced single-qubit states describe local mixedness arising from correlations with the rest of the system, the corresponding bipartitions retain the correlations responsible for this loss of purity. This perspective allows us to interpret bipartite quantities as purifying extensions of the traced partition, providing a natural bridge between local measurement properties and nonclassical correlations dictated by the symmetry and antisymmetry of the coupling structure.

With the density matrices obtained above, we are now able to analyze the system’s behavior at both the partition and bipartition levels. For single-qubit partitions, we focus on direct local measurements of the excited-state populations, which allow us to compute the corresponding Fisher information (FI) and assess the metrological sensitivity of each qubit.
For bipartitions, we investigate nonclassical properties by analyzing local Bloch vectors and linear entropy.
The linear entropy, defined as $E_{X|YZ}=1-\textrm{Tr}\{\hat{\rho}_{YZ}^{2}\}$  for a pure global state\cite{NielsenChuang}, provides a simple and intuitive quantifier of mixedness of $\hat{\rho}_{YZ}$ or $\hat{\rho}_{X}$ and bipartite entanglement between partitions $X$ and $YZ$, vanishing for pure states and increasing as correlations with the remaining subsystem build up.

\section{Local vectors}

Systems of two qubits with a Hilbert space $\mathcal{H}=\mathcal{H}_X\otimes \mathcal{H}_Y$ can be written in the Bloch representation in the following way~\cite{fano1983},
$ \hat{\varrho} = \frac{1}{4}\left(\mathbbm{1} +\vec{x}\cdot\vec\sigma\otimes \mathbbm{1} +\mathbbm{1} \otimes \vec{y}\cdot\vec\sigma + \sum_{i,j=1}^{3} t_{ij}\hat{\sigma}_i\otimes \hat{\sigma}_j\right)$,
where $\vec\sigma=(\hat{\sigma}_1,\hat{\sigma}_2,\hat{\sigma}_3)$. The \textit{local} vectors $\vec x$ and $\vec y$ can be obtained from the given bipartite quantum state $\hat{\varrho}$,
\begin{eqnarray}
    \vec x &=& \left[\textrm{Tr}(\hat{\varrho}\, \hat{\sigma}_1\otimes \mathbbm{1}), \textrm{Tr}(\hat{\varrho}\, \hat{\sigma}_2\otimes \mathbbm{1}), \textrm{Tr}(\hat{\varrho}\, \hat{\sigma}_3\otimes \mathbbm{1})\right], \nonumber \\
    \vec y &=& \left[\textrm{Tr}(\hat{\varrho}\, \mathbbm{1}\otimes \hat{\sigma}_1), \textrm{Tr}(\hat{\varrho}\, \mathbbm{1}\otimes \hat{\sigma}_2), \textrm{Tr}(\hat{\varrho}\, \mathbbm{1}\otimes \hat{\sigma}_3)\right], 
\end{eqnarray}
whereas the terms responsible for the correlation among systems can be obtained with
$t_{ij} = \textrm{Tr}[\hat{\varrho} \hat{\sigma}_i\otimes \hat{\sigma}_j]$
The elements $t_{ij}$ form a $3\times 3$ matrix, called the correlation matrix $\mathcal{T}$. In hands with the vectors $\vec x$ and $\vec y$, and the matrix $\mathcal{T}$, one can construct the matrix $\mathcal{R}$, which completely defines a general system of two-qubits and can be written as
\begin{eqnarray*}
    \mathcal{R} = \left(
    \begin{array}{cccc}
        1 & x_1 & x_2 & x_3 \\
        y_1 & t_{11} & t_{12} & t_{13} \\
        y_2 & t_{21} & t_{22} & t_{23} \\
        y_3 & t_{31} & t_{32} & t_{33}
    \end{array}
    \right) = 
    \left(
    \begin{array}{cc}
        1 & \vec x \\
        \vec{y}^T & \mathcal{T}
    \end{array}
    \right).
\end{eqnarray*}
In hand with matrix $\mathcal{R}$, it is possible to define the QO for two-qubit systems~\cite{milne2014}
\begin{eqnarray}\label{qo}
   \Omega = |\det (\mathcal{R})|^{1/4}.
\end{eqnarray}

The influence of measurements performed on one system over another system located in a distant laboratory is known as steering. This phenomenon arises only when the systems share sufficiently strong correlations. The central idea is to analyze the set of Bloch vectors to which one subsystem can steer the other, considering all possible local measurements performed on its own subsystem~\cite{jevtic2014}. These new vectors define an ellipsoidal region from which information about the correlations between the systems can be extracted. In this framework, it can be shown that the \textit{steering ellipsoid} is centered at $\vec c_X = \gamma_Y (\vec x - \mathcal{T}\vec y)$, with orientation and semiaxes lengths $s_i = \sqrt{q_i}$ given by the eigenvectors and eigenvalues $q_i$ of the ellipsoid matrix
$Q_X = \gamma_y (\mathcal{T}-\vec x \cdot \vec y^T)(\mathbbm{1}+\gamma_y \vec y\cdot \vec y^T)(\mathcal{T}-\vec y\cdot \vec x^T)$.
Here $\gamma_y = 1/(1-|\vec y|^2)$, where $|\vec y|< 1$, and we consider that measurements are being performed on the part $Y$ and the steering part $X$ of the composite system. It is important to mention this fact given that quantum steering is, by essence, non-symmetrical~\cite{uola2020}.

By using the QO~\eqref{qo}, it is possible to calculate the volume of the QSE~\cite{jevtic2014}, which gives \begin{eqnarray}\label{vqse}
    \mathcal{V}_X = \frac{4\pi}{3}\gamma_y^2 \Omega_X^4.
\end{eqnarray}
Both quantities, QO~\eqref{qo} and the volume of the QSE~\eqref{vqse}, can be used to detect quantum phase transitions in the system. The abrupt change in the derivative of these measures is a signature of such a quantum behavior~\cite{rosario2024}. 

\section{Fisher Information}

In the context of quantum parameter estimation \cite{fisher1925theory,giovannetti2011advances}, one aims to determine an unknown quantity—here, the parameter $\theta$—that characterizes the dynamics of a quantum system. When $\theta$ enters the system Hamiltonian and influences the evolution of the system's state~\cite{de2019estimation,de2022simple,de2024signal}, the theory of statistical inference provides tools to quantify how precisely $\theta$ can be estimated from experimental observations.

Assume that the system is initialized in a known quantum state and evolves under a Hamiltonian (\ref{HFull0}) $\hat{H}_{S}\rightarrow\hat{H}_{S}(\theta)$ that depends on the parameter $\theta$. After a given time evolution, a measurement is performed in the system, which yields outcomes $r$ with conditional probabilities $P_{r}(\theta)$. An estimator $\theta_{\rm est}(r)$ maps these outcomes to a numerical estimate of the parameter. For unbiased estimators satisfying the Cramér-Rao inequality, which establishes a fundamental lower limit on the precision of estimation $\delta^{2} \theta \geq \frac{1}{F(\theta)}$ \cite{scheffe1947h,rao1973linear}.

In the quantum formalism, the estimation problem is formulated along a precise line. Consider a quantum system whose state depends on an unknown parameter $\theta$ through the density operator $\hat{\rho}(\theta)$. Measurements are performed using an experimental apparatus characterized by the operator set $\{\hat\Pi_r\}$. The probability of obtaining the outcome $r$ is then given as a function of the parameter $\theta$ by $P_{r}(\theta) = {\rm Tr}[\hat\rho(\theta) \hat\Pi_r].$

A crucial point to note is that, in addition to the choice of the estimator, the uncertainty in parameter estimation also depends on the measurement strategy—in other words, it depends on the set $\{\hat\Pi_{r}\}$. The QFI is defined by maximizing the FI over all possible measurements
\begin{equation}
    {\cal F}_{Q}(\theta) =  \max_{\{\hat\Pi_{r}\}} \sum_r \frac{1}{ {\rm Tr}[\hat\rho(\theta) \hat\Pi_{r}]}\left( {\rm Tr}\left[\frac{d\hat\rho(\theta)}{d\theta} \hat\Pi_r\right]\right)^{2}.
\end{equation}

The maximization of the FI can, at least implicitly \cite{helstrom1976quantum}, be performed when a spectral representation of the density matrix is known. Consider the spectral decomposition of the density operator,
$\hat\rho(\theta)=\sum_{r}p_{r}(\theta)\vert\psi_{r}(\theta)\rangle\langle\psi_{r}(\theta)\vert$,
where the states $\vert\psi_{r}(\theta)\rangle$ form an orthonormal basis and $x$ is the parameter to be estimated. If $\hat{\rho}(\theta)$ is a pure state $|\psi(\theta)\rangle$, as is often the case in controlled qubit systems, then the QFI simplifies to
$\mathcal{F}_Q(J_{C}) = 4 \left[ \langle \partial_{\theta} \psi | \partial_{\theta} \psi \rangle - |\langle \psi | \partial_{\theta} \psi \rangle|^{2} \right]$.

This expression quantifies the distinguishability of the states $|\psi(\theta)\rangle$ with respect to infinitesimal changes in the coupling strength. A larger QFI implies that small variations in $\theta$ produce more distinguishable quantum states, and hence more precise estimations of the coupling are possible.

In the presence of decoherence or noise, the system state becomes mixed, and one must compute the QFI using the spectral decomposition
$\hat{\rho}(\theta) = \sum_n p_{n}(\theta) |\phi_n(\theta)\rangle \langle \phi_n(\theta)|$,
leading to the general expression
$\mathcal{F}_{Q}(\theta) = \sum_n \frac{[\partial_{\theta} p_n]^2}{p_n} + 2 \sum_{m\neq n} \frac{[p_m - p_n]^2}{p_m + p_n} |\langle \phi_m | \partial_{\theta} \phi_n \rangle|^{2}$,
where the first term captures classical uncertainty from population changes, and the second term arises from variations in the eigenstates \cite{braunstein1994statistical,paris2009quantum}.

Now, when we analyze the FI associated with the energy measurement in each qubit, we can treat it as classical, since the density operator of each qubit is in an incoherent state, as described in cases $\hat{\rho}_{A}=\hat{\rho}_{B}$ and $\hat{\rho}_{C}$. Thus, the probability of measuring the state $|m\rangle$, with $m = 0,1$, in each qubit $l=A,B,C$ is given by
${\cal P}_{m,l}(\theta) = \langle m \vert \hat{\rho}_{l}(\theta)  \vert m \rangle$.

The FI for this measurement is given by
\begin{equation}\label{medenerb}
F_{l}[\theta]=\sum_{m=0}^{1}\frac{\big[ \frac{d{\cal P}_{m,l}(\theta)}{d\theta} \big]^{2}}{{\cal P}_{m,l}(\theta)}.
\end{equation}

Furthermore, FI—both classical and quantum—serves as a fundamental benchmark for the sensitivity of a quantum system to variations in its coupling constants, guiding the optimal design of control and readout schemes for precise circuit characterization. Therefore, for a real measurement proposal, we will adopt in our study a local measurement, i.e., FI (\ref{medenerb}), with $\theta\rightarrow J_{C}$, which will give us an understanding of the effect of the estimation through the measurement and the filter chosen for the QO.

To establish a direct connection between local measurements and the estimation of coupling parameters, we consider the reduced state of a single qubit obtained after tracing the remaining subsystems. The resulting two-level mixed state is fully characterised by the expected values of the Pauli operators, which encode all information accessible through local observables.
Collecting these expected values in a three-dimensional real vector provides a geometric representation of the reduced state. This construction results in the definition of the FI related to the local vector in the form 
\begin{equation}\label{Fy}
F_{l}[\theta]=\frac{1}{1-\left|\vec y_{l}\right|^{2}} \left( \frac{\partial \left|\vec y_{l}\right|}{\partial \theta } \right)^{2}.
\end{equation}

Through the expression above, it becomes possible to directly compare the behavior of the local vectors with the FI, thereby revealing how local measurement sensitivity is connected to bipartite nonclassical correlations and symmetry-driven crossover phenomena. In the following, we explore this connection, together with the linear entropy, in the analysis of the quantum crossovers exhibited by the model introduced in this work.

\section{Critical Point} 

We are now in a position to investigate the model governed by the Hamiltonian~\eqref{HFull0}, with the aim of analyzing the quantum crossovers exhibited by the system. We start with the QSE/QO structures for the system under dissipative conditions. The QO~\eqref{qo} for $\hat{\rho}_{AB}$ and $\hat{\rho}_{AC}$ are given by, respectively,
\begin{eqnarray}\label{Oab}
    \Omega_{AB} = \frac{2}{\mathcal{N}}||\beta|^4-|\alpha|^2|^{1/2},
\end{eqnarray}
and
\begin{eqnarray}\label{Oac}
    \Omega_{AC} = \frac{2|\beta|}{\mathcal{N}}||\alpha|^2 - 1 |^{1/2}.
\end{eqnarray}
In order to calculate the volume of the QSE~\eqref{vqse}, we need the local vectors. For the state $\hat{\rho}_{AB}$, both local vectors (for parts $A$ and $B$) are the same and given by
\begin{eqnarray}\label{local-a}
    \vec y_{A(B)} = \left\{0,0,\frac{|\alpha|^2-1}{\mathcal{N}} \right\}.
\end{eqnarray}
For the state $\hat{\rho}_{AC}$, the local vector for part $C$ can be written as  
\begin{eqnarray}
    \vec y_{C} = \left\{0,0,-\frac{|\alpha|^2 - 2|\beta|^2 + 1}{\mathcal{N}} \right\}.
\end{eqnarray}
The local vector for part $A$ is the same as given in~\eqref{local-a}.
With these quantities, one is able to calculate the volume of the QSE for both possibilities ($\hat{\rho}_{AB}$ and $\hat{\rho}_{AC}$),
\begin{equation}\label{Vab}
    \mathcal{V}_{AB} = \frac{\pi \mathcal{N}^4}{12}\frac{(|\beta|^4-|\alpha|^2)^2}{(1+|\beta|^2)^4(|\alpha|^2+|\beta|^2)^4},
\end{equation}
and
\begin{equation}\label{Vac}
    \mathcal{V}_{AC} = \frac{\pi \mathcal{N}^4}{192}\frac{(1-|\alpha|^2)^2}{(1+|\alpha|^2)^4|\beta|^4}.
\end{equation}

\begin{figure}[h!]
\begin{center}
\includegraphics[scale=0.58,angle=0]{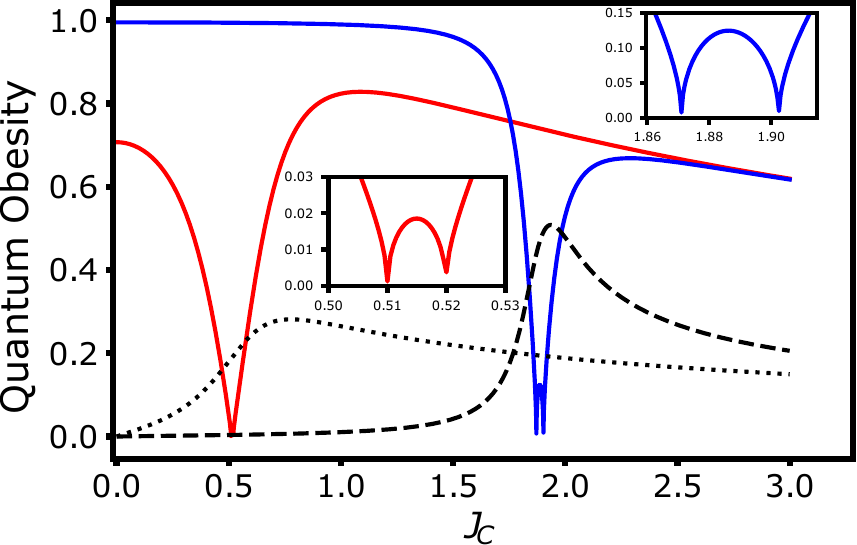} 
\vspace{1.5cm}
\includegraphics[scale=0.58,angle=0]{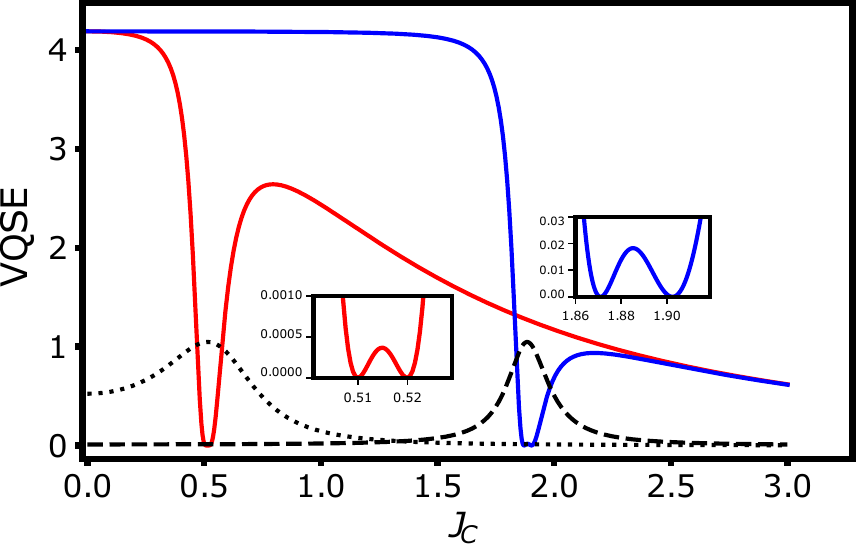}
\end{center}
\caption{(a) Quantum obesity and (b) Volume of the quantum steering ellipsoid as a function of $J_{C}$. The continuous curves represent $\Omega_{AB}$ for parameters $J=0.1$ (red curve), and $J=1$ (blue curve). The non-continuous curves represent $\Omega_{AC}$ for the parameters $J=0.1$ (dotted curve), and $J=1$ (dashed curve).} 
\label{fig-qo}
\end{figure} 

In Figs.~\ref{fig-qo}, we analyze these measures for different values of $J$. One can see that both measures agree on the values of $J_C$ to which the crossover phenomenon occurs. Note that the crossovers can also be detected when we look at the state $\hat{\rho}_{AC}$. Interestingly, one can observe a duplication of the critical points. The origin of this duplication will be discussed in the following.

\begin{figure}
\begin{center}
\includegraphics[scale=0.568,angle=0]{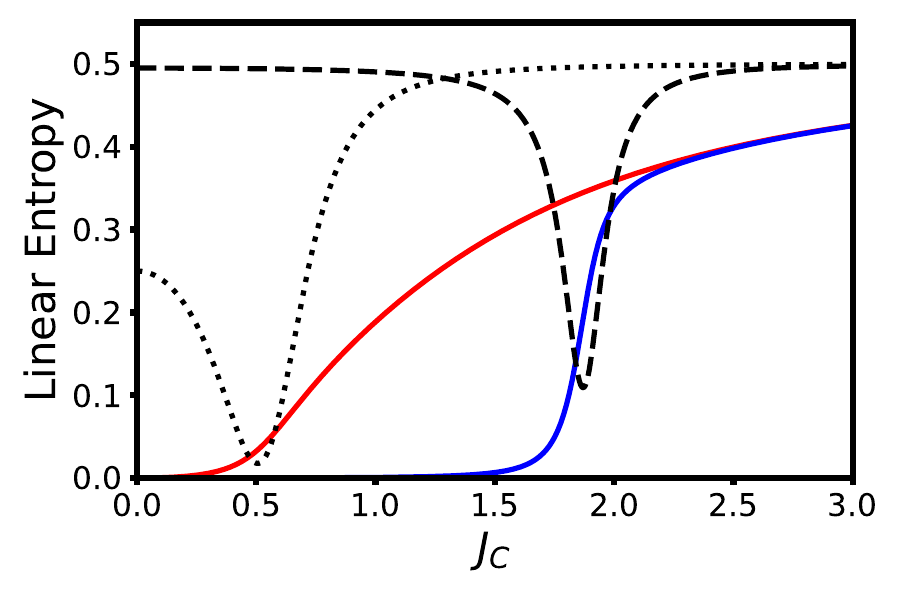}
\end{center}
\caption{Linear Entropy {\it vs} $J_{C}$, with fixed parameters $\omega_{C} = 5.0$ and $\omega_{0} = 0.1$ for all curves.  The continuous curves represent $E_{C|AB}$ for parameters $J=0.1$ (red curve), and $J=1$ (blue curve). The non-continuous curves represent $E_{B|AC}$ for the parameters $J=0.1$ (dotted curve), and $J=1$ (dashed curve).} 
\label{figle}
\end{figure}

\begin{figure}
\begin{center}
\includegraphics[scale=0.58,angle=0]{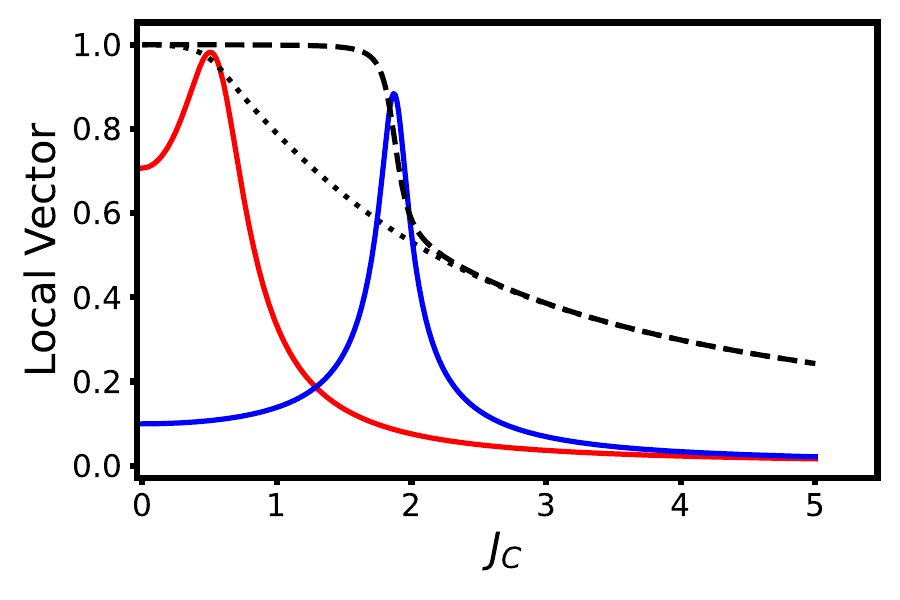}
\end{center}
\caption{Local vector {\it vs} $J_{C}$: The continuous curves represent the magnitude of the local vectors $\left|\vec{y}_{A(B)}\right|$ for parameters
 $J = 0.1$ (red curve), and $J = 1.0$ (blue curve).
The non-continuous curves represent $\left|\vec{y}_{C}\right|$ for the parameters $J = 0.1$ (dotted curve), and $J = 1$ (dashed curve).} 
\label{figLV}
\end{figure}

We now turn to the analysis of the linear entropy, shown in Fig.~\ref{figle}, which further corroborates the emergence of the crossover phenomenon. For coupling strengths below the critical value $J_C$, the partition $AB$ remains in a pure state, effectively disentangled from qubit $C$. At the critical point, a crossover takes place in which the subsystem $AB$ becomes entangled with $C$, and a pronounced reduction in the entanglement between the partitions $AC$ and $B$ is observed. Therefore, this critical process not only signals a loss of purity in the bipartite state $AB$, but also reveals a sudden reorganization of the entanglement structure, as evidenced by the crossing of the linear entropies $E_{B|AC}$ and $E_{C|AB}$. The same phenomenon can be observed between the local vectors $\left|\vec{y}_{A(B)}\right|$ and $\left|\vec{y}_{C}\right|$ in  Fig.~\ref{figLV}.

\begin{figure}
\begin{center}
\includegraphics[scale=0.58,angle=0]{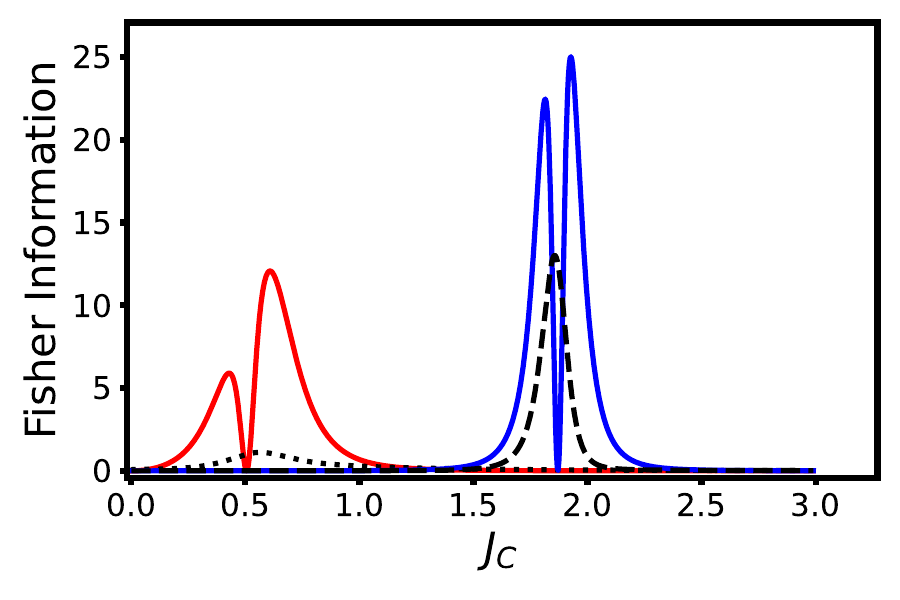}
\end{center}
\caption{Fisher information {\it vs} $J_{C}$, with fixed parameters $\omega_{C} = 5.0$ and $\omega_{0} = 0.1$ for all curves.  The continuous curves represent $F_{A(B)}$ for parameters $J=0.1$ (red curve), and $J=1$ (blue curve). The non-continuous curves represent $F_{C}$ for the parameters $J=0.1$ (dotted curve), and $J=1$ (dashed curve).} 
\label{plotFI}
\end{figure}

By analyzing the FI in conjunction with the density operators $\hat{\rho}_{A}$ and~$\hat{\rho}_{C}$, we establish a connection with the concepts of the local vector. Remarkably, this relation emerges solely from measurements of the excited state populations of each qubit, leading to the following expression (\ref{Fy}). Recalling that $F_{B}[J_{C}]=F_{A}[J_{C}]$, the expression reveals that, at critical points where the excited-state probability reaches zero $\frac{d {\cal P}_{1,A}}{d J_{C}}=\frac{d {\cal P}_{1,B}}{d J_{C}}=0$,  together with $\frac{d \vec{y}_{A} }{d J_{C}}=\frac{d \vec{y}_{B} }{d J_{C}}=0$, we observe a quantum sensitivity point characterized by $F_{A} = F_{B} = 0$, resulting in the same sensitivity effect as {\it quantum wheatstone bridge} in~\cite{poulsen2022quantum,tiwari2025quantum}. In this regime, the local measurement and the filtering operation are identical, leading to a duplication of the critical point simultaneously in both $\Omega_{AB}$ and $\mathcal{V}_{AB}$, as we can see in Fig.~\ref{fig-qo}. This behavior, however, does not occur when we analyze configurations with distinct local measurements and filtering, such as the bipartitions $AC$ or $BC$, see Fig.~\ref{plotFI}.

\begin{figure}
\begin{center}
\includegraphics[scale=1.5,angle=0]{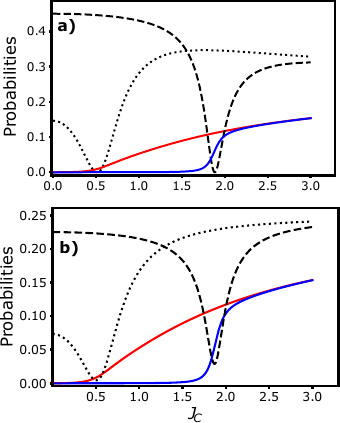}
\end{center}
\caption{Probabilities as a function of $J_{C}$. (a) The continuous curves represent $P_{010100}$ for parameters $J=0.1$ (red curve), and $J=1$ (blue curve). The non-continuous curves represent $P_{001}$ for the parameters $J=0.1$ (dotted curve), and $J=1$ (dashed curve). (b) The continuous curves represent $P_{100}$ for parameters $J=0.1$ (red curve), and $J=1$ (blue curve). The non-continuous curves represent $P_{010001}$ for the parameters $J=0.1$ (dotted curve), and $J=1$ (dashed curve).} 
\label{figp}
\end{figure}

A deeper understanding of the crossover phenomenon can be obtained by analyzing how the probability populations of the computational basis states evolve as the coupling parameter $J_C$ is varied. From Eq.~(\ref{eq:psiG}), the steady state of the system is the projector, which can be written in the computational basis $\{|000\rangle,|001\rangle,|010\rangle,|011\rangle,|100\rangle,|101\rangle,|110\rangle,|111\rangle\}$ as
\begin{equation}
|\psi_G\rangle=
\frac{1}{\sqrt{N}}
\left(
-\alpha|001\rangle
-\beta|010\rangle
-\beta|100\rangle
+|111\rangle
\right).
\end{equation}
Therefore, the probabilities of occupying each computational state are
\begin{eqnarray*}
P_{001} &=& \left| \langle \psi_{G} | 001\rangle \right|^{2} = \frac{|\alpha|^2}{N}, \\
P_{010} &=& \left| \langle \psi_{G} | 010\rangle \right|^{2} = \frac{|\beta|^2}{N}, \\
P_{100} &=& \left| \langle \psi_{G} | 100\rangle \right|^{2} = \frac{|\beta|^2}{N}, \\
P_{111} &=& \left| \langle \psi_{G} | 111\rangle \right|^{2} = \frac{1}{N}.
\end{eqnarray*}
The probabilities of occupying each superposition state are
\begin{eqnarray*}
P_{010100} &=& \left| \langle \psi_{G} | \left( \frac{|010\rangle+|100\rangle}{\sqrt{2}} \right) \right|^{2} = \frac{|\beta|^2}{N}, \\
P_{001111} &=& \left| \langle \psi_{G} | \left( \frac{|001\rangle+|111\rangle}{\sqrt{2}} \right) \right|^{2} = \frac{|\alpha|^2+1}{2N}, \\
P_{010001} &=& \left| \langle \psi_{G} | \left( \frac{|010\rangle+|001\rangle}{\sqrt{2}} \right) \right|^{2} = \frac{|\alpha|^2+|\beta|^2}{2N}.
\end{eqnarray*}
The population inversions can be seen in Figs.~\ref{figp}. 

By comparing the probability curves, it becomes possible to gain a deeper understanding of the emergence of crossovers in the system, as evidenced by the QFI, linear entropy, and local vectors. In this way, one can identify that the origin of the crossover is associated with an inversion of probabilities induced by the system coupling. Furthermore, the QO and VQSE measures do not predict the crossover at the same critical points, reinforcing the conclusion that they are not reliable indicators of this phenomenon. In particular, the crossing pattern observed in Fig. \ref{figp}.a shows that the crossover can be revealed through local measurements performed on qubit $A$ or $B$, since the corresponding probabilities exhibit the same inversion behavior associated with the critical point. On the other hand, the behavior displayed in Fig.\ref{figp}.b is qualitatively different: the probabilities cross in an opposite manner, which does not allow the crossover to be detected through a local measurement on qubit $C$.

\section{Discussion}

In conclusion, we have shown that quantum crossover phenomena in few-body open systems can be faithfully captured through purely local measurements, establishing a direct connection between local Fisher information and the emergence of crossover behavior. Beyond this correspondence, we demonstrate that quantum obesity does not, in general, extend the quantum steering ellipsoid volume as a universal indicator of crossovers. In particular, we identify regimes in which the QSE volume fails to signal the transition, while the underlying physics remains encoded in the behavior of the local Bloch vector. This distinction arises from the fact that the parameter $\gamma_y$ is not constant across the crossover, implying that the characterization of critical behavior is governed primarily by the local vector structure rather than by the global ellipsoidal volume. Finally, in the Supplementary information, we provide a detailed blueprint for demonstrating crossover indicators with tunable couplings, establishing a concrete and experimentally relevant framework that supports and extends the results presented in this work. Our results clarify the geometric origin of crossover signatures and highlight the fundamental role of local state properties in diagnosing nontrivial quantum behavior in few-body systems.

\appendix
\section{Solution to the master equation}

Our analysis is based on the knowledge of the global ground state of the system described in Eq.~\eqref{HFull0} that reduces the steady state of the system~\eqref{eq:PMEspin0} . The ground state for the full Hamiltonian (\ref{HFull0}) can be analytically written as~\cite{diniz2021quantum} 
\begin{eqnarray*}
    \left| \psi_G \right\rangle &=& \frac{1}{\sqrt{\mathcal{N}}}
    \begin{pmatrix}
 0, &
- \alpha, &
 - \beta, &
  0, &
 - \beta, &
   0, &
   0, &
   1
\end{pmatrix}^{T},
\end{eqnarray*}
with the respective quantities defined by 
\begin{eqnarray*}
    \mathcal{N} = 1+|\alpha|^2 +2|\beta|^{2} ,
\end{eqnarray*}
\begin{eqnarray*}
    \alpha = \frac{-8J_C^2-4J\omega - 2J\omega_C - 2\omega\omega_C-\omega_C^2 -2t J + 2\omega t + t^2}{2(2J^2-4J_C^2+J\omega_C-Jt)} ,
\end{eqnarray*}
\begin{eqnarray*}
    \beta = \frac{2 J J_C+ 2J_C\omega + J_C \omega_C + J_C t}{2J^2-4J_C^2+J\omega_C-Jt}.
\end{eqnarray*}
Here,
\begin{eqnarray*}
    t &=& \frac{1}{3}\Big\{2J-\omega_C-\frac{A\, (1+i\sqrt{3})}{(B+\sqrt{-A^3+B^2})^{1/3}}\\
    &+&(-1+i\sqrt{3})(B+\sqrt{-A^3+B^2})^{1/3}\Big\} ,
\end{eqnarray*}
\begin{eqnarray*}
    A &=& 4J^2 + 12 J_C^2 + 3\omega^2 + 2J \omega_C + \omega_C^2, \nonumber \\
    B &=& -8J^3 + 72 J J_C^2 - 9J\omega^2 - 6J^2 \omega_C + 18 J_C^2 \omega_C \\
    &-& 9 \omega^2 \omega_C + 3J \omega_C^2 + \omega_C^3.
\end{eqnarray*}

From the ground state given above, we are able to construct the density operator for the full system, i.e., $\hat{\rho}_{SS}  = |\psi_G\rangle\langle\psi_G|$ and the reduced matrices, which are given by the following
\begin{eqnarray*}
\hat{\rho}_{AB} &=& \frac{1}{\mathcal{N}}   
\left(
\begin{array}{cccc}
|\alpha|^2 & 0 & 0 & -\alpha \\
0 & |\beta|^2 & |\beta|^2 & 0 \\
0 & |\beta|^2 & |\beta|^2 & 0 \\
-\alpha^{\ast} & 0 & 0 & 1
\end{array}
\right),
\end{eqnarray*} 

\begin{eqnarray*}
\hat{\rho}_{AC} = \hat{\rho}_{BC} &=& \frac{1}{\mathcal{N}}   
\left(
\begin{array}{cccc}
|\beta|^2 & 0 & 0 & -\beta \\
0 & |\alpha|^2 & \alpha\beta^{\ast} & 0 \\
0 & \alpha^{\ast}\beta & |\beta|^2 & 0 \\
-\beta^{\ast} & 0 & 0 & 1
\end{array}
\right),
\end{eqnarray*} 

\begin{eqnarray*}\label{rhoa}
\hat{\rho}_{A} &=& \hat{\rho}_{B} = \frac{1}{\mathcal{N}}   
\left(
\begin{array}{cc}
|\alpha|^2 + |\beta|^2 & 0 \\
0 & 1 + |\beta|^2
\end{array}
\right),
\end{eqnarray*} 

\begin{eqnarray*}\label{rhoc}
\hat{\rho}_{C} &=& \frac{1}{\mathcal{N}}   
\left(
\begin{array}{cc}
2|\beta|^2 & 0 \\
0 & 1 + |\alpha|^2
\end{array}
\right).
\end{eqnarray*}

\section{Blueprint for demonstrating crossover indicators with tunable couplings}

The coherent coupling of quantum bits (qubits) is one of the foundational elements that enable quantum technologies, particularly in quantum computation, simulation, and metrology~\cite{montenegro2025quantum,paris2009quantum}. Among the main types of couplings implemented in superconducting circuits~\cite{niskanen2007quantum,riwar2022circuit,de2011generation}, there exist capacitive and inductive couplings tunable. Capacitive coupling~\cite{moskalenko2022high,yan2018tunable,de2025interferometric} arises from shared electric fields between neighboring qubits — often transmons — through a common capacitance, enabling direct exchange of excitations. Inductive coupling~\cite{neill2018blueprint}, on the other hand, leverages mutual inductance between superconducting loops, where interactions are mediated by shared magnetic flux~\cite{van2005mediated,campbell2023modular}. Each coupling strategy brings specific advantages in terms of tunability, coherence, and integration~\cite{huang2018universal,bialczak2011fast}. Hybrid strategies combining both capacitive and inductive elements have been proposed to engineer more complex and controllable interaction topologies~\cite{chen2014qubit,moskalenko2021tunable,heunisch2023tunable}.

These coupling architectures are central to the scalability of quantum processors~\cite{field2024modular,sarma2025designing,egorova2024high,sete2021floating,arute2019quantum}. A landmark demonstration is the work by Google’s Quantum AI team, which introduced a programmable superconducting quantum processor in their influential work~\cite{neill2018blueprint}. In this work, the authors proposed a two-dimensional lattice of qubits with tunable nearest-neighbor couplings that laid the groundwork for later experiments, including the celebrated 9-qubit processor. These systems employ dynamically tunable couplings via external flux control, enabling reconfigurable Hamiltonians that are critical for both quantum supremacy demonstrations and quantum simulation tasks.

Theoretical models for such architectures quickly become intractable even when limited to a few qubits. For closed quantum systems (i.e., without dissipation), it is still possible to derive analytical solutions for systems with up to four qubits under symmetry constraints~\cite{dos2024four}. However, when considering open quantum systems, where interactions with an external environment are described by master equations, the situation becomes more complex. Non-unitary evolution introduces decoherence and dissipation channels that entangle the system with its surroundings, making exact analytical solutions difficult even for systems with as few as three qubits~\cite{diniz2021quantum}. This reflects a fundamental challenge in simulating realistic and experimentally relevant quantum circuits.

\begin{figure}[h!]
\begin{center}
\includegraphics[scale=0.081,angle=0]{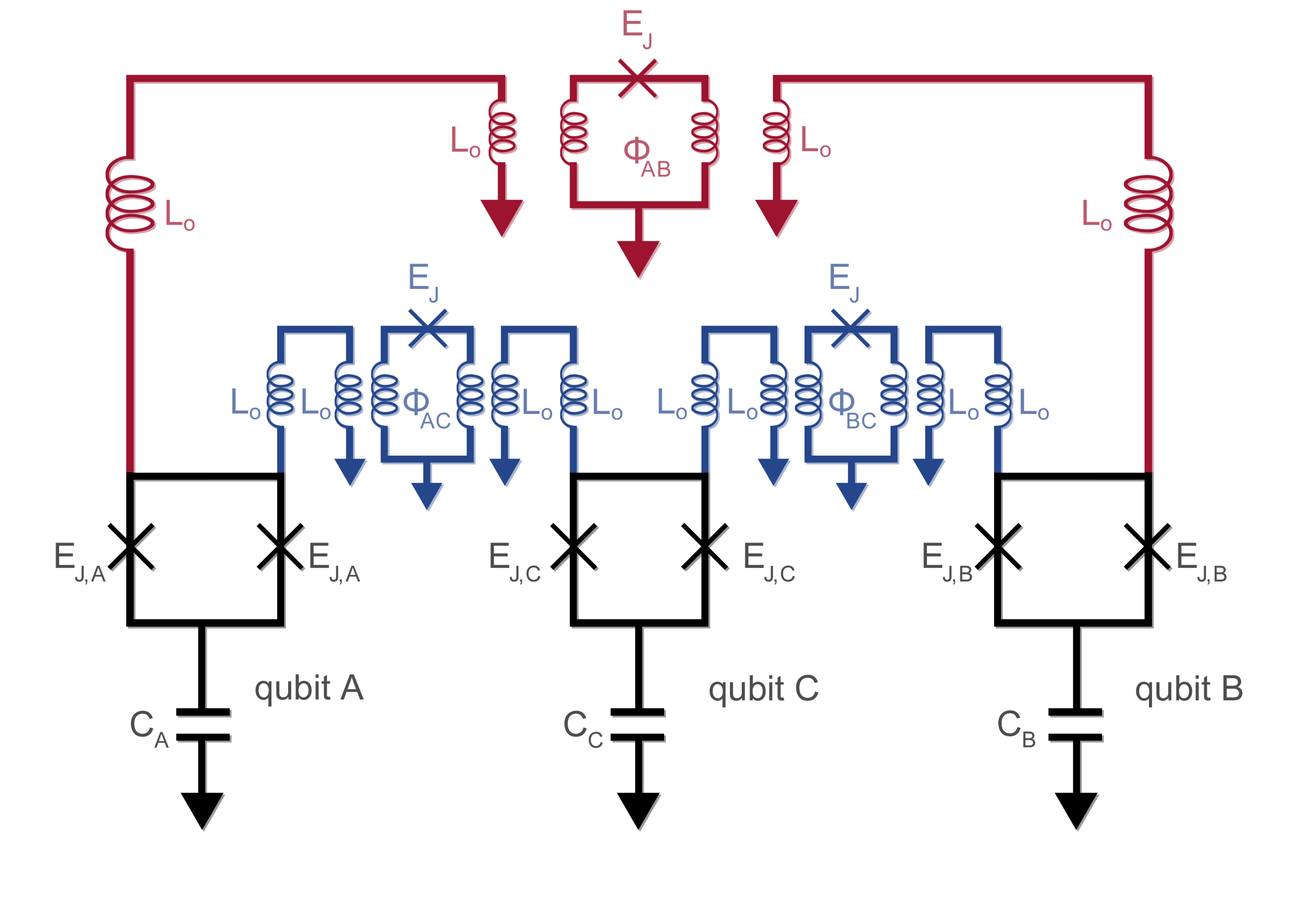}
\end{center}
\caption{Schematic diagram showing the circuit implementation in a three-qubit inductive tunable coupler.} 
\label{fig0}
\end{figure}

We consider a superconducting quantum device composed of three transmon qubits, labeled qubit $A$, $B$, and $C$. Each pair of qubits is connected through a tunable inductive element, such as a symmetric superconducting quantum interference device (SQUID), which enables flux-controlled coupling. This configuration allows independent modulation of the interaction strength between each pair---($A,B$), ($A,C$), and ($B,C$)---via externally applied magnetic fluxes $\Phi_{AB}$, $\Phi_{AC}$, and $\Phi_{BC}$, respectively (see figure \ref{fig0}).

Each transmon qubit consists of a Josephson junction with Josephson energy $E_{J,i}$ shunted by a capacitance $C_i$, leading to a charging energy $E_{C,i} = e^{2}/(2C_i)$. In the transmon regime ($E_{J,i} \gg E_{C,i}$), and the Hamiltonian of each isolated qubit is given by
\begin{equation}
\hat{H}_{j} = 4E_{C,j} \hat{n}_{j}^2 - E_{J,j} \cos( \hat{\varphi}_{j}), \quad (j = A, B, C),
\end{equation}
where $\hat{\varphi}_{j}$ is the superconducting phase across the Josephson junction and $\hat{n}_{j}$ is the corresponding Cooper-pair number operator~\cite{de2025exploring}, with $[ \hat{\varphi}_{j},\hat{n}_{j} ]=i$.

The interaction between any two qubits $j$ and $k$ is mediated by a tunable inductive element whose effective inductance depends on an externally applied flux $\Phi_{jk}$. The corresponding interaction Hamiltonian is expressed as
\begin{equation}
    H_{jk}^{\text{int}} = \frac{1}{2L_{ij}(\Phi_{jk})} \left( \Phi_j - \Phi_k \right)^2  \rightarrow \hat{H}_{jk}^{\text{int}} = \frac{E_L^{(jk)}(\Phi_{jk})}{2} (\hat{\varphi}_j - \hat{\varphi}_k)^2,  
\end{equation}
where $\Phi_j \rightarrow (\Phi_0 / 2\pi) \hat{\varphi}_j$ is the flux variable associated with qubit $j$, $\Phi_0 = h / (2e)$ is the flux quantum, and $E_L^{(jk)}(\Phi_{jk})$ is the flux-dependent inductive energy defined as
\begin{equation}
E_L^{(jk)}(\Phi_{jk}) = \left( \frac{\Phi_0}{2\pi} \right)^2 \frac{1}{L_{jk}(\Phi_{jk})}.
\end{equation}
The effective inductance of the SQUID follows
\begin{equation}
L_{jk}(\Phi_{jk}) = \frac{L_0}{|\cos(\pi \Phi_{jk} / \Phi_0)|},
\end{equation}
where \(L_0\) is the minimum inductance when $\Phi_{jk} = 0$.

The full Hamiltonian of the system, including the three transmon and all pairwise interactions ($\cos(\hat{\varphi}_{j})\approx-\hat{\varphi}_{j}/2$), is
\begin{eqnarray}
\hat{H}_{t} &=& \sum_{j=A,B,C} \left[ 4E_{C,j} \hat{n}_{j}^2 + \frac{E_{J,j}}{2} \hat{\varphi}_{j}^{2} \right] \nonumber \\
&&+ \sum_{j\neq k} \frac{E_L^{(jk)}(\Phi_{jk})}{2} (\hat{\varphi}_{j} - \hat{\varphi}_{k})^2.    
\end{eqnarray}
To obtain an effective transmon-level description, we quantize the phase and charge operators using
\begin{equation}
\hat{\varphi}_{j} = \varphi_{\text{ZPF},i} (\hat{a}_{j} + \hat{a}_{j}^{\dag}), \quad \hat{n}_{j} = n_{\text{ZPF},i} \frac{(\hat{a}_{j}-\hat{a}_{j}^{\dag} )}{i},
\end{equation}
where the zero-point fluctuation (ZPF) amplitudes are
\begin{equation}
\varphi_{\text{ZPF},j} = \left( \frac{2E_{C,j}}{E_{J,j}} \right)^{1/4}, \quad n_{\text{ZPF},j} = \left( \frac{E_{J,j}}{32E_{C,j}} \right)^{1/4}.
\end{equation}

Projecting the Hamiltonian onto the lowest two energy levels of each transmon, $\hat{a}_{j}\rightarrow\hat{\sigma}_{-}^{j}=\left|0\right\rangle\left\langle1\right|^{j}$, we obtain the effective Hamiltonian in the qubit subspace:
\begin{eqnarray}
\hat{H}_{\text{eff}} &=& \hbar \sum_{i=A,B,C} \frac{\omega_i}{2} \hat{\sigma}_z^{i} + \hbar g_{AB}(\Phi_{AB}) \hat{\sigma}_x^{A} \hat{\sigma}_x^{B}  \nonumber\\
&& + \hbar g_{AC}(\Phi_{AC}) \hat{\sigma}_{x}^{A} \hat{\sigma}_x^{C} + \hbar  g_{BC}(\Phi_{BC}) \hat{\sigma}_{x}^{B} \hat{\sigma}_x^{C},
\label{Heff}
\end{eqnarray}
where $\hat{\sigma}_{x}^{j}$, $\hat{\sigma}_{y}^{j}$, and $\hat{\sigma}_{z}^{j}$ are the Pauli matrices,
\begin{eqnarray*}
\hbar \omega_{j} &=& \sqrt{8E_{C,j}E_{J,j}} + \sum_{j\neq k} \frac{E_{L}^{(jk)}(\Phi_{jk})}{2} \approx \sqrt{8E_{C,j}E_{J,j} } ,
\end{eqnarray*}
is the transition frequency of qubit $j$, and the flux-tunable coupling rates are given by
\begin{equation}
\hbar g_{jk}(\Phi_{ij}) \equiv E_{L}^{(jk)}(\Phi_{jk}) \varphi_{\text{ZPF},j} \varphi_{\text{ZPF},k}.
\end{equation}

This architecture enables real-time control of the interaction topology by dynamically modulating the external fluxes $\Phi_{jk}$. Specifically, each coupling term $g_{jk}(\Phi_{jk})$ can be turned on, off, or modulated in strength, allowing the system to emulate different coupling geometries (linear, triangular, or star-like) and to implement multi-qubit gate operations or quantum simulations with high flexibility.

\subsection*{Acknowledgments} ACSC thank CNPq-Brazil under Grant N$^{o}$ 308730/2023-2. ECD acknowledge the UNEMAT (Campus Tangará da Serra) for the hospitality. ECD also thanks Gabriel T. Landi for discussions at the beginning of this work on the numerical calculations of Quantum Fisher Information and Agrientech Tecnologia Agrícola $\&$ Ambiental for the computational resources provided in the initial stages of this work. OPSN is partially supported by  FAPEPI, via Edital N$^{o}$ 004/2025: {\it ``Programa de Bolsas de Produtividade em Pesquisa (PPQ) da UESPI''}.

\bibliography{apssamp1}

\end{document}